\documentstyle[11pt,aaspp4,flushrt]{article}

\newcommand{\pomega}{\mbox{$\varpi$}}
\newcommand{\erfc}{\mbox{\rm erfc}}

\begin{document}

\title{The Globular Cluster Luminosity Function as a Distance Indicator: \\
       Dynamical Effects}

\author{Jeremiah P. Ostriker \& Oleg Y. Gnedin}

\affil{Princeton University Observatory, 
       Peyton Hall, Princeton, NJ~08544;
       jpo,ognedin@astro.princeton.edu}

\begin{abstract}
The dynamical evolution of the globular cluster systems in galaxies
is predicted, based on the standard dynamical theory normalized to
the example of the Milky Way. The major processes varying with the
galactocentric distance are the tidal shocks and dynamical friction.
Our simple model explains, on a quantitative basis, the observed
differences of the inner and outer populations of globular clusters.
We can thus calculate corrections for dynamical evolution for the
luminosity function of globular clusters with the assumption
that the initial luminosity function is identical in all galaxies
(and we can test this assumption as well, in certain cases).
Then we can compute the expected distribution of absolute magnitudes
and compare it with the observed distribution of apparent magnitudes
to estimate the distance moduli for M31 and M87. Using this new method
we find ($dm_{\rm M31}=24.05\pm 0.23, \; dm_{\rm M87}=30.83\pm 0.17$)
as compared to current best estimates using other methods of
($dm_{\rm M31}=24.30\pm 0.20, \; dm_{\rm M87}=31.0\pm 0.1$).
As a check on the method we compute, and compare with observations,
the differences between the inner and outer globular clusters in
all three galaxies. This new method, coupled with HST observations,
promises to provide an independent method of estimating distances to
galaxies with recession velocities $\lesssim 10,000$ km s$^{-1}$,
or $d \lesssim 100h^{-1}$ Mpc.
\end{abstract}


\section{Introduction}
Evolution of the globular cluster systems (GCS) in our Galaxy and external
galaxies has received a burst of recent interest. New HST observations of
the GCS in the Virgo ellipticals allow a direct comparison of the globular
cluster luminosity functions (GCLF) in different galaxies. The assumption
that the peak of the GCLF occurs at a constant luminosity for galaxies
of a given type, and varies slowly and in a predictable manner with the
galactic type, has been used widely as a distance indicator
(\cite{H:91}; \cite{Jetal:92}; \cite{H:96}) although the theoretical basis
for this expectation remains unclear.

On the theoretical front, \cite{GO:97} and \cite{MW:96} (1996a,b,c)
extending earlier work of \cite{AHO:88} and \cite{CW:90} have
shown that significant evolution of the GCS is to be expected over
the Hubble time. The inner populations of globular clusters are most
affected by the tidal shocks and dynamical friction. The first of these
effects discriminates preferentially against the less dense and less
massive clusters, the latter affecting the most massive clusters, to shape
the currently observed distribution of luminosities. We have shown
(Gnedin 1997) that the expected statistical ``brightening'' of the inner
clusters (due, really, to the destruction of some of the less luminous
clusters) is indeed observed in the three best studied galaxies, the Milky Way,
M31, and M87 (see also \cite{vdB:96}). In this paper we examine the
hypothesis that a common initial distribution of clusters existed in the
inner and outer parts of all the studied galaxies with the apparent
variations due to the influence of dynamical evolution. And then we show
how the method can be used to obtain an independent estimate of the distance
for galaxies with an observed distribution of globular cluster apparent
magnitudes.

\section{The mass -- density distribution in the Milky Way}
The primary processes driving globular cluster evolution are
two-body relaxation, evaporation of stars through the tidal limit,
tidal shocks, and dynamical friction. Relaxation is the only internal
effect and in zeroth approximation does not depend on the position
of the cluster in the galaxy. The tidal effects do depend on the density
of the galactic environment and tend to destroy less dense clusters. The rate
of inspiral due to dynamical friction is proportional to the cluster mass,
$M$, thus preferentially destroying the most massive clusters.
Both the tidal effects and dynamical friction are strongly enhanced
in the inner part of the galaxy and are weak on the outskirts. Therefore
it is possible that the outer populations of globular clusters may still bear
the form of the initial distribution, whereas the inner population
is affected by the extrinsic dynamical effects. Passive evolution,
through stellar interior processes and two-body relaxation, is expected
for all clusters. A more careful treatment would allow for the interaction
between internal and external effects.

First we shall make an attempt to
reconstruct the original distribution using the outer cluster sample.
The simplest generalization of the GCLF is a two-dimensional distribution
of the cluster mass and density. A useful estimate of the cluster
density is $\rho \equiv M/R_h^3$, where $R_h$ is the half-mass radius.
Since only the Galactic clusters are well resolved, we use the sample of
the Milky Way clusters tabulated by \cite{D:93}. The core-collapsed
clusters were eliminated from the sample due to the observational difficulty
of estimating their radii. We also rejected the few very distant clusters
(beyond 60 kpc from the Galactic center) as being of uncertain origin.
The remaining sample consists of 94 clusters. We set the boundary between
the two samples at the galactocentric radius $R_{1/2}=5$ kpc, thus putting
36 clusters in the inner and 58 clusters in the outer groups. The outer
sample has more members because we expect the inner population to be
significantly depleted. This choice also maximizes the difference between
the two populations within the range $4<R_{1/2}<7$ kpc.

\begin{figure}
\plotone{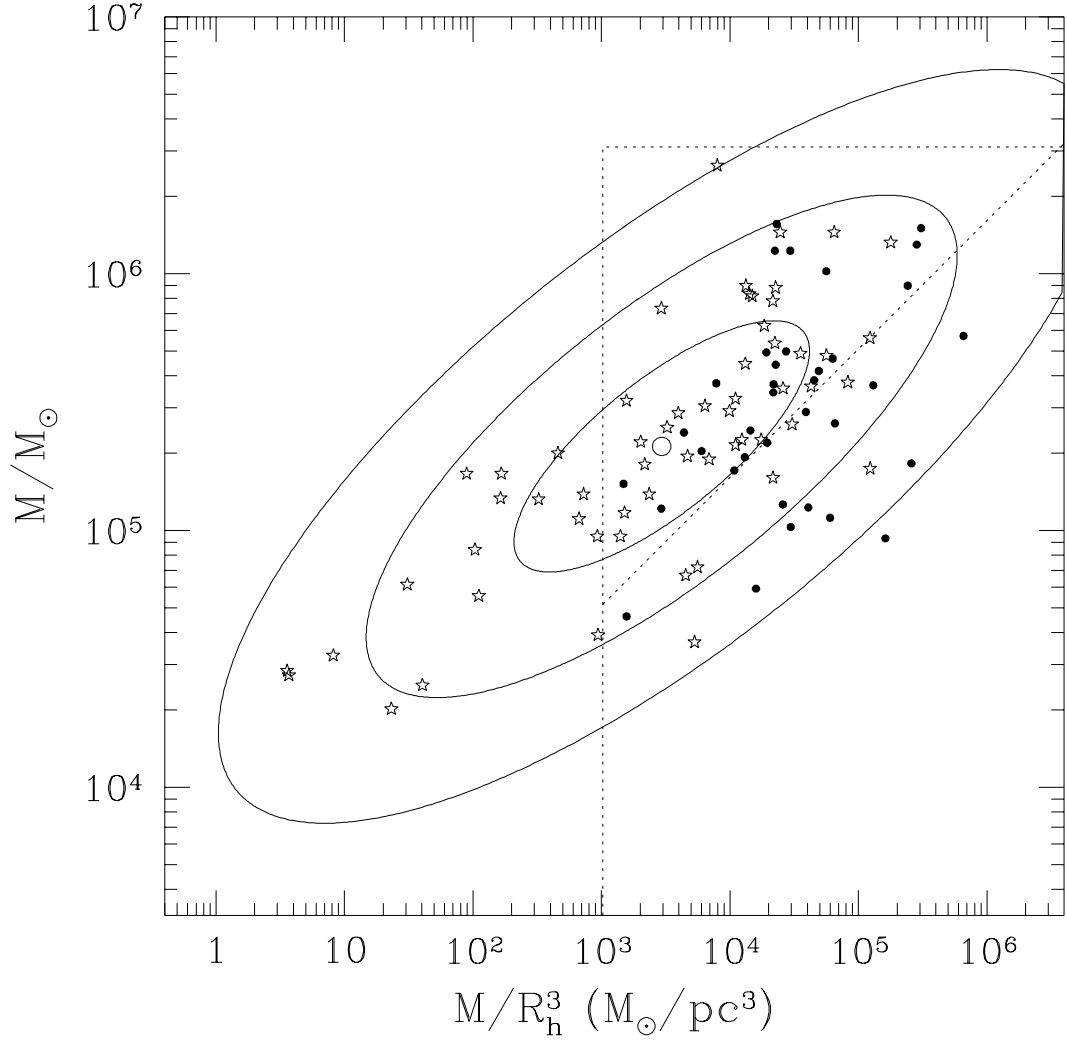}
\caption{Distribution of the inner (filled dots) and outer (stars)
Galactic globular clusters on the $\log{M}-\log{\rho}$ plane.
The solid lines show the adopted initial distribution function
(eq. [\protect\ref{eq:df}]) where the ellipses are $1\sigma$, $2\sigma$,
and $3\sigma$ levels, and the small circle is at the center
of the distribution. The dotted lines mark the region allowed for the
inner clusters by dynamical processes: above the horizontal line
dynamical friction is important, to the left of the vertical line tidal
shocks are important, and below the diagonal line relaxation will lead
to core collapse and subsequent disintegration of the clusters. The
equivalent of the first two of the survival boundaries for the outer
clusters are not shown as they do not significantly constrain cluster
properties.
\label{fig:massden}}
\end{figure}

Figure \ref{fig:massden} shows the distribution of the defined sample of
Galactic globular clusters in the $\log{M} - \log{\rho}$ plane.
Throughout, we use an assumed mass-to-visual-light ratio of $M/L_V = 3$
to estimate masses when velocity dispersions are not available.
The area occupied by the inner clusters is visibly constrained by the
survival boundaries of dynamical friction and the tidal shocks.
The boundaries for the inner clusters (the horizontal and vertical dotted
lines) are defined as the median clusters that would marginally survive
in the age of the Galaxy conservatively estimated as $t_G = 12$ Gyr.
For details see \cite{GO:97}.

As noted by \cite{AHO:88} and \cite{GO:97}, dynamical friction
does not pose a serious cutoff on the distribution of the clusters
in Milky Way. It will, however, be seen to be important for the M87
globulars because their distribution function is much better sampled
(see \S \ref{sec:m87}). The tidal shocks are, on the contrary, very
important in destroying the low density clusters.

Notice a very strong correlation between the mass and density of the outer
clusters. The Pearson correlation coefficient is 0.76 and the Spearman
coefficient is 0.77, so the probability of no correlation is less than
$10^{-16}$. The two-dimensional least-squares fit gives a useful relation
between the half-mass radius and mass of globular clusters that has not
been investigated before:
\begin{equation}
R_h = \left( {2\times 10^6\, M_{\sun} \over M} \right)^{0.63} \; {\rm pc},
\end{equation}
or approximately $R_h \propto M^{-2/3}$. It is worth noting that this
relation predicts a very steep dependence on mass of the tidal shock
destruction time, $t_{sh} \propto M R_h^{-3} \propto M^3$, and a shallow
dependence on mass of the relaxation time,
$t_{rh} \propto M^{1/2} R_h^{3/2} \propto M^{-1/2}$.
As to whether or not this correlation is primordial remains to be seen,
since the present sample could have been affected by relaxation itself.

Including the correlation in the bivariate distribution
reduces the occupied space on the $\log{M}-\log{\rho}$ plane and
maximizes the distribution function $f(\log{M},\log{\rho})$. While the
log--luminosity distribution is known to be approximately Gaussian, the
distribution of $\log{\rho}$ is certainly skewed. Nevertheless, we assume
the bivariate Gaussian shape for the combined distribution for the sake
of tractability. Thus the adopted {\it primordial} distribution function
of globular cluster properties is taken to be
\begin{equation}
f(x,y) \, dx dy = {a + 1/a \over 2\pi \sigma_1 \sigma_2} \, \exp{\left[
   -{(y-ax-b)^2 \over 2\sigma_1^2} -{(y+x/a-c)^2 \over 2\sigma_2^2} \right]} \,
   dx dy,
\label{eq:df}
\end{equation}
where $x \equiv \log{\rho}, \, y \equiv \log{M}; \, a = 0.35$ is the slope of
the correlation, and the constants $b$ and $c$ are fixed by the center of the
two-dimensional distribution. The dispersion along the line of the correlation,
$\sigma_2=3.70$, is much larger than that in the perpendicular direction,
$\sigma_1=0.32$. The units adopted are such that the cluster mass is in
solar masses and the density is in $M_{\sun}$ pc$^{-3}$. The solid lines
in Figure \ref{fig:massden} show the adopted distribution function. It
provides a satisfactory fit to the outer clusters (denoted by stars).

Given the initial distribution, we can evolve it in time and compare with the
present data for the inner population. The most straightforward approach
is to assume a hard cutoff at the density $\rho_{sh}$ and mass $M_{df}$
such that the clusters less dense or more massive than that would have been
destroyed within the Hubble time by the tidal shocks or dynamical friction,
respectively. Such survival boundaries are shown on Figure \ref{fig:massden}.
Then, since for most observed GCSs we do not know individual values of
$R_h$ and hence $\rho$, we integrate the evolved distribution function
over $\rho$ (hence $x$) to obtain the evolved mass distribution which
can be used to compute the luminosity distributions when studying the
GCSs in external galaxies. The result of the integration is
\begin{eqnarray}
F(y) \, dy = {1 \over \sqrt{2\pi} \sigma_y} \, e^{-(y-y_0)^2/2\sigma_y^2} \;
  \theta(\mu-y) \; {1 \over 2} \erfc(\pomega) \, dy, \label{eq:df2} \\
  \pomega(y) \equiv \sqrt{\alpha} \varrho + {y-c \over 2a\sigma_2^2
  \sqrt{\alpha}} - {a(y-b) \over 2\sigma_1^2 \sqrt{\alpha}}, \hspace{.5cm}
  \alpha \equiv {a^2 \over 2\sigma_1^2} + {1 \over 2a^2 \sigma_2^2}, \nonumber
\end{eqnarray}
where $\mu \equiv \log{M_{df}}$, $\varrho \equiv \log{\rho_{sh}}$,
$\theta(\mu-y)$ is a step function, and \erfc(\pomega)\ is the complementary
error function. Equation (\ref{eq:df2}) should also be renormalized to
the size of the remaining sample. An important consequence of the modified
distribution is the shift of the peak, given by the nonlinear equation
\begin{equation}
\Delta y \equiv y - y_0 = - \left( {d\pomega \over dy} \right) \sigma_y^2 \;
  {2 e^{-\pomega^2} \over \sqrt{\pi} \, \erfc(\pomega)},
\label{eq:dy}
\end{equation}
where we neglected the dynamical friction correction as it is usually not
important for the bulk of the sample. This equation can be easily solved
iteratively to obtain $\Delta y$. When all parameters of the initial
distribution (eq. [\ref{eq:df}]) are fixed, the predicted shift is a
function only of $\varrho$, the amplitude of the tidal shocks. It was
calculated approximately, equating the inverse destruction rate of each
cluster due to the shocks, $t_{sh}$ (see Gnedin \& Ostriker 1997),
to the Hubble time.
A correction factor of order unity can be used to calibrate $\varrho$
against the data. Taking $t_{sh}/0.3$ as true destruction time gives
the observed shift between the inner and outer clusters in Milky Way
($\Delta y \approx 0.16$), and we adopt this factor here. The correction
in magnitudes is related to the shift $\Delta y$ by
$\Delta m_0 = -2.5 \Delta y$.
The dispersion of the inner sample can also be calculated similarly.

\begin{figure}
\plotone{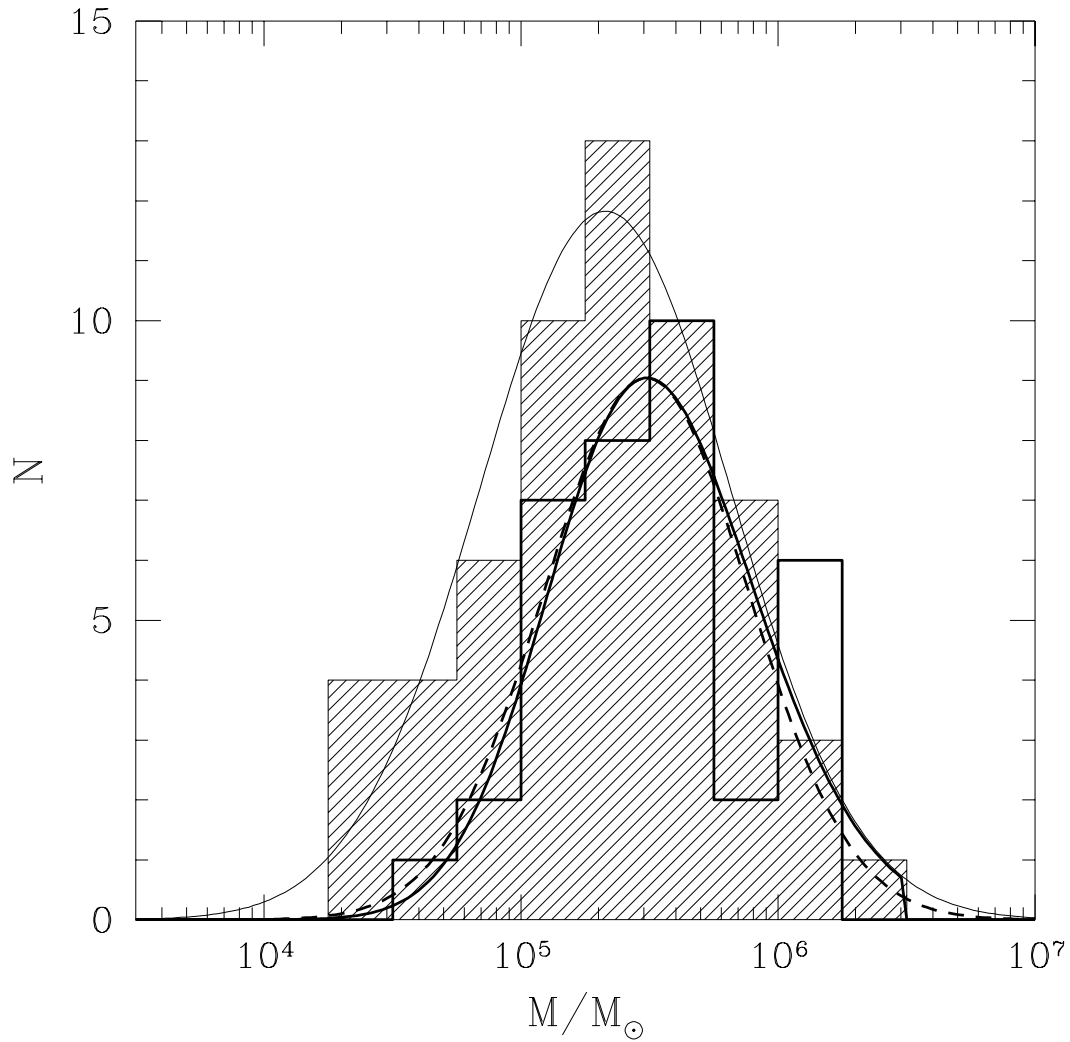}
\caption{The histograms of the inner (solid line) and outer (shaded)
samples of the Galactic globular clusters. The thick dashed line and the
thin solid line are the Gaussian fits to the inner and outer populations,
respectively. The thick solid line is the predicted evolved
distribution (eq. [\protect\ref{eq:df2}]) for the inner clusters
assuming that they started with the same properties as the outer clusters
but suffered from dynamical evolution.
\label{fig:mw2}}
\end{figure}

The observed and predicted mass distributions of the Galactic globular clusters
are shown on Figure \ref{fig:mw2}. The agreement is excellent.
Our correction reproduces well both the mean and the dispersion of
the inner sample. The first of these two checks is of course not
significant (but only a check of consistency) since we fixed the factor
0.3 by the requirement that the observed and computed inner peaks agree
at $M_{V,in}^0 = -7.73$. But the correct predicted reduction of the dispersion
$\Delta \sigma = 0.25\pm 0.12$ mag seen between the two samples
($\Delta \sigma_{obs} = 0.23\pm 0.16$) provides a fair test of the hypothesis
that the inner population had the same initial properties as the outer
population but was modified by dynamical effects.
Table \ref{tab:deltas} provides a comparison of the predicted and observed
changes in the mean and the dispersion, as well as the associated uncertainty.

\section{Predicted evolution in M31 and M87}
\label{sec:m87}
The predicted evolution of the mass (and luminosity) function of globular
clusters can now be immediately applied to the samples in M31 and M87
to test if the predicted values of both $\Delta m_0$ and $\Delta \sigma$
match observations for these independent samples. We use the same data as
Gnedin (1997) who gives the appropriate references.

The amplitude of the tidal shocks scales with the density
of the galaxy at a given position, which is
$\rho \propto M_{gal}(R)/R^3 \propto v_c^2/R^2$
for a spherically symmetric system. Here $v_c$ is the rotation speed
(or the velocity dispersion for elliptical galaxies)
at the median radius $R_{med}$ of the globular clusters from the center
of the galaxy.
While the velocity can be directly measured, only angular distances are
available for external galaxies so the physical distance will scale
with the assumed distance to the galaxy, $R_{med} \propto D$.
The dynamical friction cutoff also depends on both $v_c$ and $R_{med}$
through $M_{df} \propto v_c R_{med}^2$, where we neglect the slowly
varying logarithmic factor.
Even though both $\rho_{sh}$ and $M_{df}$ depend on the distance $D$
to the galaxy, the correction (eq. [\ref{eq:dy}]) can be applied in a
self-consistent manner without any initial assumptions about the
distance. The method proposed below is fully independent and will
{\it predict}, rather than use, the distance $D$.

If the observed samples is rich enough it could be divided into
inner and outer halves, or even further; the method can be applied to each
of the subpopulations separately and the distance estimated as average
over the individual results. In principle, the narrower the division
of the sample into radial bins the better the accuracy of the median
estimator $R_{med}$; in reality, the subsamples should have a large
enough number of members to represent well the luminosity function.
For M31 and M87 we take the inner and outer halves of the sample,
as defined by Gnedin (1997).
For M31, the inner clusters have the median radius at
$R_{in,med}=13.\arcmin5$, whereas the outer clusters have
$R_{out,med}=50\arcmin$. The rotation velocity of M31 is estimated to be
$v_c \approx 300$ km s$^{-1}$ for $10\arcmin < R < 20\arcmin$, and
$v_c \approx 330$ km s$^{-1}$ for $R > 20\arcmin$ (\cite{Fetal:90}).

Let us demostrate how the method works on the example on the inner
population of globular clusters in the Andromeda. The mean magnitude
of the subsample $m_{0,in}=15.93$ (Gnedin 1997) so the trial value
of the distance modulus is taken to be $dm_{(1)} = m_{0,in}-M_V^0=23.26$, where
$M_V^0 = -7.33$ is the center of the assumed initial distribution
of globular clusters. Using this distance we calculate the critical
shock density $\rho_{sh,\rm M31} = 5.4\times \rho_{sh,\rm MW}$ and the
corresponding correction to the peak magnitude, $\Delta m_{0,in}=0.80$.
The peak of the inner GCLF therefore becomes dimmer and the second
estimate of the distance modulus is $dm_{(2)}=24.06$. Now second
and further iterations are performed to calculate new $\rho_{sh}$ and
to obtain new estimate of $dm$ until convergence is achieved. Usually
4 or 5 iterations are sufficient to obtain the distance modulus to an
accuracy better than 0.01 mag. In case of the inner clusters of M31,
the final predicted shift of the peak magnitude is $\Delta m_{0,in}=0.65$
and the distance is $dm_{\rm M31}=23.91$. The same operations are then
repeated for the outer clusters as well, starting with their mean
$m_{0,out}=16.71$. The results are given in Table \ref{tab:deltas}.
Now combining the two predicted changes for the inner and outer populations
we get an estimate of $\Delta m_0$ and $\Delta \sigma$ for that galaxy.
The predicted change agrees with the observed values within the errors
(Table \ref{tab:deltas}).

The distance modulus of M31 has been obtained by a number of other methods.
The Cepheid variables give $dm=24.43\pm 0.16$ (\cite{FM:90}),
RR Lyrae give $dm=24.34\pm 0.15$ (\cite{PvdB:87}), fitting color-magnitude
diagrams of population II giants gives $dm=24.40\pm 0.25$ (\cite{MK:86}),
and novae give $dm=24.04\pm 0.20$ (\cite{C:85}). The average of all these
methods indicate $dm_{\rm M31}=24.30\pm 0.20$. Our method predicts
$dm_{\rm M31}=24.05\pm 0.23$ (averaged over the inner and outer samples;
see Table \ref{tab:distances}), in good agreement with the other methods.
Note that our result was derived independently from any other
distance determinations to the galaxy.

In case of the Virgo giant M87 the dispersion of radial velocities $v_c$
is used instead of the rotation velocity as an estimate of the galaxy mass.
\cite{MT:93} give the mass of M87 of $6\times 10^{12}\, M_{\sun}$
at $R=50$ kpc, so that $v_c \approx 730$ km s$^{-1}$.
The globular clusters are distributed over a larger area; the median
radius of the inner population is at $R_{in,med} = 2.\arcmin4$,
and of the outer population is at $R_{out,med} = 5.\arcmin1$.
Repeating the same steps as for the M31 sample, we obtain a very good
agreement of the predicted changes with the observed
$\Delta m_0 = 0.26\pm 0.03$ and $\Delta \sigma = 0.04\pm 0.03$
(see Table \ref{tab:deltas}). The distance modulus to M87
is essentially the same as inferred from the inner or outer samples
and is $dm_{\rm M87} = 30.83\pm 0.17$, in remarkably good agreement with
the value of $dm_{\rm M87} = 31.0\pm 0.1$ as obtained for the center of
the Virgo cluster based on Cepheids, planetary nebulae luminosity function,
and surface brightness fluctuations (quoted by \cite{H:96}).
The largest source of error in the method proposed here is due to the
uncertainty in the peak of the Galactic globular clusters,
$\delta M_V^0 = 0.16$ mag.

Because of the extreme richness of the clusters in M87, the effects of
dynamical friction should be detectable. Indeed, if we keep the same
division of the clusters into the inner and outer populations but consider
only the brightest tails of both, with $V < 20.9$, the outer clusters
become relatively brighter (as the brightest inner clusters are depleted).
This effect is illustrated in Table \ref{tab:m87}.
Statistically significant, it shows lack of very
bright, and presumably massive, clusters in the inner region, as is
expected, if they spiraled into the center by dynamical friction.
The medians provide an even stronger case at the $5\sigma$ level.

Clusters at $R = 12$ kpc would reach the center of M87 in the Hubble
time due to dynamical friction if they are more massive than
$1.4\times 10^7\, M_{\sun}$ (\cite{BT:87}, p. 	428).
This corresponds to the visual magnitude of $V_{df} = 19.3$. 
In accord, only 4 clusters brighter than $V_{df}$ are present in the inner
sample while the outer sample has 9 as bright members. As suggested by
\cite{OBS:89}, \cite{Petal:92}, and \cite{CD:93}, the dynamical friction rate
is strongly enhanced in triaxial halos with box orbits; this effect can
lower even further the mass cutoff. Therefore dynamical friction is
important in determining distribution of very bright clusters in M87.

\section{Discussion}
By applying the corrections for dynamical evolution, the GCLF can be
used as an improved distance indicator. For example, the method can be
used for determining distances to the Virgo cluster and, possibly, to the
Coma cluster. HST offers
the unprecedented opportunity to detect faint globular cluster systems
around galaxies in the Virgo cluster. Having a large sample of such
galaxies is essential to eliminate another source of error,
the difference in metallicity of different galaxies (see \cite{ACZ:95}).

We have hitherto assumed a fixed mass-to-light ratio whereas it is
known that both the luminosity of individual stars of fixed mass and
the distribution of masses within clusters are a function of the cluster
mean metallicity. An {\it ab initio} correction for such effects would
be difficult and uncertain. But an empirical approach which utilizes the
fact that for an evolved population, the mean color is a good proxy for
metallicity, should be possible. The minimization of the effects due to
metallicity differences could be
done as follows. Assume that the ($B-V$) color of the integrated galaxy
light is a fair
indicator of the metallicity. Then the mean, $m_{0,i}$, of the GCLF of
each galaxy $i$ is assigned a metallicity correction of the form
$\delta m_i = \alpha \, \delta(B-V)_i \equiv \alpha [(B-V)_i -
 \overline{(B-V)}]$,
where $\alpha$ is a constant to be determined,
in addition to the correction $\Delta m_{dyn,i}$ for dynamical evolution;
here $\overline{(B-V)}$ is the mean color of the Galactic sample.
The true mean point is thus
\begin{equation}
\overline{m}_i = m_{0,i} + \Delta m_{dyn,i} + \alpha \, \delta(B-V)_i.
\end{equation}
We now consider a sample of galaxies all approximately at the same distance
(for example, in the Virgo cluster) but with different mean colors
and metallicities. Then
\begin{equation}
\langle m \rangle = \langle m_{0,i} + \Delta m_{dyn,i} \rangle
  + \alpha\, \langle \delta(B-V)_i \rangle.
\label{eq:mav}
\end{equation}
We then minimize the dispersion of the individual means $\overline{m}_i$ around
the average $\langle m \rangle$, to determine $\alpha$:
\begin{equation}
\alpha = {\langle m_{0,i}+\Delta m_{dyn,i} \rangle\langle \delta(B-V)_i \rangle
  - \langle (m_{0,i}+\Delta m_{dyn,i}) \delta(B-V)_i \rangle \over
  \langle \delta(B-V)_i^2 \rangle - \langle \delta(B-V)_i \rangle^2}.
\end{equation}
Using this value of $\alpha$, we obtain a statistical estimator
(eq. [\ref{eq:mav}]) of the
peak of the GCLF to be calibrated against the known absolute magnitude,
$M_V^0$, of the GCLF in the Milky Way and M31. The proposed corrected
estimator has the advantage of being justified by simple, relatively well
understood physics. When used in a real survey of globular clusters
in external galaxies, it may significantly reduce the scatter among
the results for different galaxies.

\section{Conclusions}
The differences in the inner and outer populations of globular clusters
in our own Galaxy (Gnedin 1997) can be explained in terms of simple
dynamical processes such as the tidal shocks and dynamical friction.
Assuming that the primordial distribution of globular clusters is common
for all galaxies, we derive the correction to the turnover point of the GCLF
in any galaxy based on the dynamical calculations for the Galactic clusters
by \cite{GO:97}. The predicted shifts between the mean points of the inner
and outer cluster populations agree qualitatively with the observed
differences within the errors. The shifts in the dispersion of luminosity
between the inner and outer samples of the Galaxy, M31, and M87 are
predicted accurately. A slight mismatch of the mean points
for the Milky Way and M31 of the order 0.2 mag still remains, but is within
the statistical errors of the methods applied.
 
The corrections for dynamical evolution thus allow one to use globular
cluster systems in external galaxies as an independent distance indicator.
For a large sample of galaxies at the same distance, as in the Virgo
and Coma clusters, the metallicity corrections could also be used with
the aim of reducing still further the scatter in the derived distances.
An extensive survey of the GCS in distant galaxies, performed with the HST,
may help produce a reliable distance ladder on the scale of a hundred Mpc.

\acknowledgements
This project was supported in part by NSF grant AST 94-24416.

\pagebreak
\begin{deluxetable}{lccc}
\tablecaption{Predicted and observed differences in the luminosity function
              \label{tab:deltas}}
\tablecolumns{4}
\tablehead{& \colhead{Milky Way} & \colhead{M31} & \colhead{M87}}
\startdata
$\Delta m_{0,predicted}$ (mag)
                        & $0.41\pm 0.16$ & $0.51\pm 0.18$ & $0.25\pm 0.17$ \nl
$\Delta m_{0,observed}$ (mag)
                        & $0.40\pm 0.23$ & $0.78\pm 0.12$ & $0.26\pm 0.03$ \nl
\tableline
$\Delta \sigma_{predicted}$ (mag)
                        & $0.25\pm 0.12$ & $0.16\pm 0.12$ & $0.07\pm 0.12$ \nl
$\Delta \sigma_{observed}$ (mag)
                        & $0.23\pm 0.16$ & $0.24\pm 0.08$ & $0.04\pm 0.03$
\enddata
\end{deluxetable}

\begin{deluxetable}{lcc}
\tablecaption{Distance moduli to M31 and M87 \label{tab:distances}}
\tablecolumns{3}
\tablehead{& \colhead{M31} & \colhead{M87}}
\startdata
This paper    & $24.05 \pm 0.23$ & $30.83 \pm 0.17$ \nl
Other methods & $24.30 \pm 0.20$ & $31.0 \pm 0.1$
\enddata
\end{deluxetable}

\begin{deluxetable}{lcccc}
\tablecaption{The bright tail of the globular cluster distribution in M87
              \label{tab:m87}}
\tablecolumns{5}
\tablehead{\colhead{Sample} & \colhead{N} & \colhead{$m_0$ (mag)} &
           \colhead{$\sigma$ (mag)} & \colhead{$\mu$ (mag)}}
\startdata
Inner & 108 & $20.82\pm 0.05$ & $0.63\pm 0.03$ & $20.56\pm 0.03$ \nl
Outer &  93 & $20.75\pm 0.07$ & $0.85\pm 0.05$ & $20.37\pm 0.10$ \nl
difference  & & $-0.07\pm 0.03$ &              & $-0.20\pm 0.04$ \nl
probability & & $3.1\times 10^{-2}$ &          & $1.2\times 10^{-5}$
\enddata
\tablecomments{Clusters with $V<20.9$ and $1.'21<R<7'$. $N$ is the number of
clusters in the sample, and $m_0$, $\sigma$, and $\mu$ are the mean,
dispersion, and the median, respectively. The data is from
\protect\cite{MHH:94}. Probability quoted is the probability of Monte-Carlo
simulations that a given result occurs by chance.}
\end{deluxetable}


\begin{thebibliography}{}
\bibitem[Aguilar, Hut, \& Ostriker (1988)]{AHO:88}
Aguilar, L., Hut, P., \& Ostriker, J. P. 1988, \apj, 335, 720

\bibitem[Ashman, Conti \& Zepf 1995]{ACZ:95}
Ashman, K. M., Conti, A., \& Zepf, S. E. 1995, \aj, 110, 1164

\bibitem[Binney \& Tremaine 1987]{BT:87}
Binney, J., \& Tremaine, S. 1987, Galactic Dynamics (Princeton:
Princeton Univ. Press)

\bibitem[Capuzzo-Dolcetta (1993)]{CD:93}
Capuzzo-Dolcetta, R. 1993, \apj, 415, 616

\bibitem[Chernoff \& Weinberg (1990)]{CW:90}
Chernoff, D. F., \& Weinberg, M. D. 1990, \apj, 351, 121

\bibitem[Cohen 1985]{C:85}
Cohen, J. G. 1985, \apj, 292, 90

\bibitem[Djorgovski (1993)]{D:93}
Djorgovski, S. 1993, in Structure and Dynamics of Globular Clusters,
ASP Conf. Ser., vol. 50, ed. S. G. Djorgovski, \& G. Meylan 
(San Francisco: ASP), p. 325

\bibitem[Federici, Fusi Pecci \& Marano 1990]{Fetal:90}
Federici, L., Fusi Pecci, F., \& Marano, B. 1990, \aap, 236, 99

\bibitem[Freedman \& Madore 1990]{FM:90}
Freedman, W. L., \& Madore, B. F. 1990, \apj, 365, 186

\bibitem[]{}
Gnedin, O. Y. 1997, \apjl, submitted; preprint {\tt astro-ph/9610171}

\bibitem[Gnedin \& Ostriker (1997)]{GO:97}
Gnedin, O. Y., \& Ostriker, J. P. 1997, \apj, 474 (in print)

\bibitem[Harris 1991]{H:91}
Harris, W. E. 1991, \araa, 29, 543

\bibitem[Harris 1996]{H:96}
Harris, W. E. 1996, in The Extragalactic Distance Scale,
   eds. M. Donahue, M. Livio (STScI, Baltimore)

\bibitem[Jacoby et al. 1992]{Jetal:92}
Jacoby, G. H., et al. 1992, \pasp, 104, 599

\bibitem[McLaughlin, Harris, \& Hanes (1994)]{MHH:94}
McLaughlin, D. E., Harris, W. E., \& Hanes, D. A. 1994, \apj, 422, 486

\bibitem[Merritt \& Tremblay (1993)]{MT:93}
Merritt, D., \& Tremblay, B. 1993, \aj, 106, 2229

\bibitem[Mould \& Kristian 1986]{MK:86}
Mould, J., \& Kristian, J. 1986, \apj, 305, 591

\bibitem[Murali \& Weinberg]{MW:96}
Murali, C., \& Weinberg, M. D. 1996a, preprint {\tt astro-ph/9602058}

\bibitem[Murali \& Weinberg 1996b]{MW:96b}
Murali, C., \& Weinberg, M. D. 1996b, preprint {\tt astro-ph/9604049}

\bibitem[Murali \& Weinberg 1996c]{MW:96c}
Murali, C., \& Weinberg, M. D. 1996c, preprint {\tt astro-ph/9610229}

\bibitem[Ostriker, Binney \& Saha (1989)]{OBS:89}
Ostriker, J. P., Binney, J., \& Saha, P. 1989, \mnras, 241, 849

\bibitem[Pesce, Capuzzo-Dolcetta, \& Vietri (1992)]{Petal:92}
Pesce, E., Capuzzo-Dolcetta, R., \& Vietri, M. 1992, \mnras, 254, 466

\bibitem[Pritchet \& van den Bergh 1987]{PvdB:87}
Pritchet, C. J., \& van den Bergh, S. 1987, \apj, 316, 517

\bibitem[van den Bergh 1996]{vdB:96}
van den Bergh, S. 1996, preprint {\tt astro-ph/9610165}

\end{thebibliography}
\end{document}